\documentclass[a4paper,11pt]{article}

\usepackage{jcappub}
\usepackage[T1]{fontenc}
\usepackage{natbib}
\usepackage[normalem]{ulem}
\usepackage{amsmath}
\usepackage{amssymb}
\usepackage{geometry}
\usepackage{booktabs}
\usepackage{array}
\usepackage{paralist}
\usepackage{verbatim}
\usepackage{graphicx}
\usepackage[utf8]{inputenc}
\usepackage{textcomp}
\usepackage{multirow}
\usepackage[dvipsnames*,svgnames]{xcolor}
\usepackage{wasysym}
\usepackage{subfigure}
\usepackage{enumitem}
\usepackage{lineno}
\usepackage{float}
\usepackage{gensymb}

\title{\textbf{Probing Dark Photon Dark Matter with CTAO}}
\author{Júlia G. Mamprim,}
\emailAdd{juliagouveamamprim@usp.br}

\author{Aion Viana and}
\emailAdd{aion.viana@ifsc.usp.br}

\author{Vitor de Souza}
\emailAdd{vitor@ifsc.usp.br}

\affiliation{Instituto de F\'isica de S\~ao Carlos, Universidade de S\~ao Paulo, Av. Trabalhador S\~ao-carlense 400, S\~ao Carlos, Brazil}

\abstract{The dark photon is a hypothetical gauge boson arising in extensions of the Standard Model, and has emerged as a compelling dark matter candidate. As dark photon dark matter (DPDM), it can interact with electromagnetic fields via kinetic mixing, and the inelastic scattering process $\gamma A^\prime \to e^+ e^-$ becomes kinematically allowed for gamma rays above a characteristic energy threshold. This interaction imprints unique spectral attenuation features at very-high-energies (VHE), offering an observational probe of DPDM models. Using the Cherenkov Telescope Array Observatory (CTAO) Instrument Response Functions (IRFs), we simulate observations of VHE sources and forecast novel sensitivities to the kinetic mixing parameter for the photon-dark photon scattering process. Our study focuses on three key astrophysical targets: the Crab Nebula and the blazars Markarian 421 and Markarian 501. Additionally, we investigate the impact of dark matter spikes around black holes on the upper limits. For the dark matter spike scenario considered in this work, CTAO observations could probe the DPDM parameter space down to a mixing parameter of $\varepsilon \sim 10^{-8}$ for masses around $m_{A^\prime}\sim 10^{-1}\,\mathrm{eV}$ through high-energy spectral attenuation, at a 95\% confidence level.}

\begin{document}

\maketitle
\flushbottom

\section{Introduction}

Dark Matter (DM) constitutes approximately 27\% of the total energy density of the Universe and dominates the matter content on cosmological scales~\cite{Planck:2018vyg,Persic:1995ru}. Its existence is firmly supported by a wealth of astrophysical and cosmological evidence, including galaxy rotation curves~\cite{1970ApJ...159..379R}, cluster dynamics~\cite{Clowe:2006eq}, the Cosmic Microwave Background (CMB)~\cite{Planck:2018jri}, and large-scale structure formation~\cite{SDSS:2005xqv}. Despite this strong astrophysical evidence, the particle nature of DM remains unknown, with no confirmed signals from direct detection experiments (e.g., LUX~\cite{LUX:2017ree}, XENONnT~\cite{XENON:2023cxc}, PandaX-4T~\cite{PandaX:2022aac}) or indirect searches (e.g., Fermi-LAT~\cite{Fermi-LAT:2015att}, H.E.S.S.~\cite{HESS:2013rld}, IceCube~\cite{IceCube:2023ies})(see e.g. \cite{Cirelli:2024ssz} for a comprehensive summary of constraints across different channels). 

Hidden sector extensions of the Standard Model have emerged as compelling frameworks to address the nature of dark matter. Among these models, one possibility involves the introduction of a new Abelian gauge boson, the dark photon $A^\prime$, associated with an additional $U(1)'$ symmetry~\cite{HOLDOM1986196,FAYET1990743}. In this scenario, dark photons produced non-thermally in the early Universe can account for all or a fraction of the dark matter content observed today~\cite{Nelson:2011sf,PaolaArias_2012}. Through kinetic mixing with the SM photon, dark photon dark matter (DPDM) leads to observable effects, despite being otherwise extremely weakly coupled~\cite{Redondo:2008aa, Fabbrichesi:2020wbt}.

A potential phenomenological signature of DPDM is its inelastic interaction with high-energy photons through the process $\gamma A^\prime \to e^+ e^-$, which becomes kinematically allowed when the center-of-mass energy exceeds twice the electron mass. For dark photon masses in the eV range, this mechanism can lead to energy-dependent attenuation features in the spectra of VHE gamma-ray sources~\cite{Bernal:2022xyi,PhysRevD.103.043018}. Such spectral signatures may serve as probes of DPDM models in observations conducted with present and next-generation gamma-ray telescopes.

In this work, we use the most recent Cherenkov Telescope Array Observatory (CTAO) Instrument Response Functions (IRFs) to simulate observations of three well-known VHE sources: the Crab Nebula, and the blazars Markarian~421 and Markarian~501. The main contributions of this study are: (i) a dedicated CTAO sensitivity forecast for photon-dark photon scattering signatures induced by dark photon dark matter in VHE gamma-ray spectra, using realistic CTAO IRFs together with a full forward-folding spectral likelihood analysis based on Poisson statistics at the count level; and (ii) a quantitative assessment of the improvement that CTAO observations could provide over current VHE constraints on the kinetic mixing parameter. By adopting benchmark source and astrophysical assumptions aligned with previous VHE studies, our comparison is designed to isolate the impact of CTAO's improved instrumental performance. We find that CTAO could strengthen the sensitivity to the kinetic mixing parameter by up to one to two orders of magnitude in the most favorable mass range. We analyze the expected attenuation features in the gamma-ray spectra induced by the inelastic scattering process to derive projected constraints on $\varepsilon$ over a range of dark photon masses. Additionally, we examine how the presence of dark matter overdensities, or ``spikes'', around supermassive black holes can enhance the predicted attenuation from blazars.

This paper is structured as follows. In Section~\ref{sec:scattering}, we summarize the theoretical framework of photon–dark photon scattering and present the relevant cross-section for this process. Section~\ref{sec:density} discusses the dark matter density profiles adopted in our analysis, including standard Navarro–Frenk–White profiles and dark matter spikes around supermassive black holes. Section~\ref{sec:sources} provides an overview of the astrophysical sources considered (Crab Nebula, Mrk 421, and Mrk 501), highlighting their spectral energy distributions and the associated dark matter environments. Section~\ref{sec:simulations} details the simulation procedure using CTAO IRFs and the modeling of the gamma-ray spectra for each source. In Section~\ref{sec:fit}, we describe the statistical methodology used to fit the models and derive projected upper limits  on the kinetic mixing parameter. Our results are presented in Section~\ref{sec:results}, followed by concluding remarks in Section~\ref{sec:conclusion}.

\section{{Photon-dark photon scattering}}
\label{sec:scattering}

The phenomenology of the massive dark photon is characterized by its interaction with Standard Model (SM) particles, as described by the Lagrangian \cite{Fabbrichesi:2020wbt}

\begin{equation}
\mathcal{L} \supset -\frac{e\varepsilon}{\sqrt{1-\varepsilon^2}} J_{\mu}A^{\prime \mu} \simeq -e\varepsilon J_{\mu}A^{\prime\mu},
\label{eq:darkphoton}
\end{equation}
where $A^{\prime \mu}$ is the gauge field associated with the dark photon, $\varepsilon$ is a dimensionless mixing parameter, $e$ is the electromagnetic coupling constant, and $J_\mu$ represents the SM electromagnetic current.

The low-energy effective theory emerges when considering energy scales $E$ satisfying the hierarchy $m_{A^\prime} \lesssim E \ll m_f$, where $m_f$ denotes the mass scale of hypothetical heavy fermions charged under both the Standard Model 
\(U(1)_{\text{EM}}\) and the hidden 
\(U(1)^\prime\) gauge groups, which mediate the interaction between the visible and dark sectors. In this regime, integrating out the heavy fermions yields the effective Lagrangian

\begin{equation}
\mathcal{L} \supset 
-\frac{1}{4} \mathcal{F}^{\mu\nu} \mathcal{F}_{\mu\nu} - \frac{\varepsilon}{2} \mathcal{F}^{\mu\nu} \mathcal{F}'_{\mu\nu} - \frac{1}{4} \mathcal{F}'^{\mu\nu} \mathcal{F}'_{\mu\nu},
\end{equation}
where $F_{\mu\nu}$ and $F^\prime_{\mu\nu}$ are the field strength tensors for the SM photon and dark photon, respectively. 

\begin{figure}[t]
    \centering
    \includegraphics[width=0.8\textwidth]{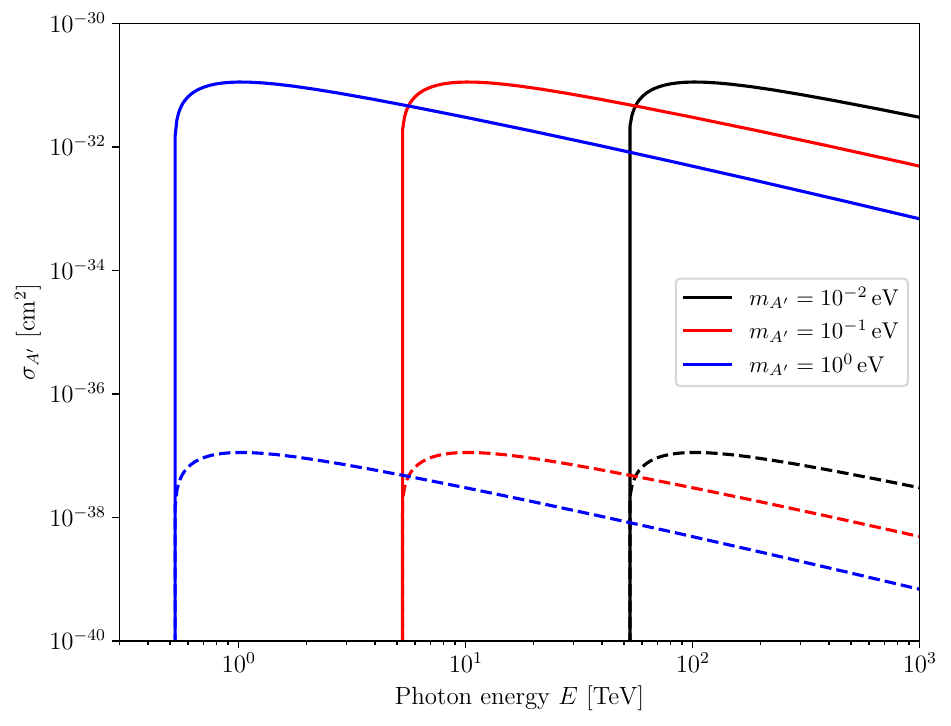}
    \caption{
    Photon–dark-photon scattering cross section $\sigma_{A'}$ as a function of the photon energy for representative dark-photon masses.
    Solid (dashed) lines correspond to $\varepsilon = 10^{-3}$ ($10^{-6}$).
    }
    \label{fig:xsec}
\end{figure}

If the DP has a non-zero mass, the scattering process $\gamma A^\prime \to e^+ e^-$ becomes kinematically accessible for gamma rays with energies above the threshold

\begin{equation}
E > E_{\text{th}}' = \frac{2m_e^2}{m_{A^\prime}},
\label{eq:threshold}
\end{equation}
where $m_e$ is the mass of the electron and $m_{A^\prime}$ is the mass of the dark photon. At leading order, the photon can undergo scattering with the two transversely polarized modes of the dark photon, mediated by its coupling to electrons through mixing with the QED photon. The corresponding scattering cross-section is~\cite{PhysRevD.103.043018}

\begin{equation}
\begin{split}
\sigma_{A^\prime} =
\frac{8\pi \varepsilon^2 \alpha^2}
{3\left(s-m_{A^\prime}^2\right)^3}
\Bigg[
&-\beta \left(s^2 + 4s m_e^2 + m_{A^\prime}^4\right) \\
&+ \ln\left(\frac{1+\beta}{1-\beta}\right)
\left(
s^2 + 4s m_e^2 + m_{A^\prime}^4
-4m_{A^\prime}^2 m_e^2
-8m_e^4
\right)
\Bigg],
\end{split}
\label{eq:xsec}
\end{equation}
with $\beta = \sqrt{1 - 4m_e^2/s}$ and $s = 2E m_{A^\prime} + m_{A^\prime}^2$. The dependence of the scattering cross section on the photon energy is shown in Figure~\ref{fig:xsec}.

This process can produce a detectable attenuation effect in the spectrum of VHE gamma ray sources. Assuming the dark photon constitutes all the dark matter in the Universe, the photon survival probability induced by the photon-dark photon scattering is 

\begin{equation}
P_{\text{sur}} = \text{e}^{-\tau_{A^\prime}},
\label{eq:survival}
\end{equation}
where $\tau_{A^\prime}$ is the optical depth due to the DPDM scattering, expressed as the integral along the line of sight (LOS) of the DPDM number density $n_{A^\prime} = \rho_\text{DM}/m_{A^\prime}$

\begin{equation}
\tau_{A^\prime} = \int_{\text{los}}n_{A^\prime}\sigma_{A^\prime}dl.
\label{eq:tau_general}
\end{equation}

Note that the survival probability depends on the dark matter density profile $\rho_\text{DM}$, and this dependence varies  with the location of the source and the properties of the intervening medium along the line of sight. 

\section{Dark matter density profiles}
\label{sec:density}

The distribution of dark matter in galaxies and galaxy clusters is commonly described using spherically symmetric halo density profiles, $\rho_{\text{DM}}(r)$. These profiles can vary significantly depending on the morphology, mass, and formation history of the system. Therefore, for the different astrophysical sources considered in this study, distinct dark matter density profiles must be adopted to accurately represent the underlying dark matter distributions.

The Zhao~\cite{Zhao:1995cp} profile provides a unified framework for characterizing dark matter spherical halos through a power-law function

\begin{equation}
\rho_\mathrm{Zhao}(r) = \rho_s \left(\frac{r}{r_s}\right)^{-\gamma}
\left[1 + \left(\frac{r}{r_s}\right)^\alpha\right]^{(\gamma-\beta)/\alpha},
\label{eq:zhao}
\end{equation}
where the characteristic density $\rho_s$ and scale radius $r_s$ set the overall normalization and size of the halo, respectively. The inner slope parameter $\gamma$ governs the central cusp behavior ($\rho \propto r^{-\gamma}$), while $\beta$ controls the outer slope ($\rho \propto r^{-\beta}$). The transition parameter $\alpha$ determines how sharply the profile changes between these regimes.

\subsection{Navarro-Frenk-White}

 Emerging from N-body simulations of collisionless dark matter, the Navarro-Frenk-White (NFW)~\cite{Navarro:1996gj} profile represents a canonical halo density distribution of dark matter in $\Lambda$CDM cosmology. It emerges by choosing $(\alpha, \beta, \gamma) = (1, 3, 1)$ in the Zhao profile (Eq.~\ref{eq:zhao}). Explicitly, 
 
\begin{equation}
    \rho_{\mathrm{NFW}}(r) = \frac{\rho_s}{( r/r_s)(1+r/r_s)^2}.
    \label{eq:NFW}
\end{equation}
For the Milky Way (MW), we assume $r_s = 14.46~\rm{kpc} $ and $\rho_s = 0.566~\rm{GeV/cm^3}$~\cite{Cirelli:2024ssz}, constrained by the local dark matter density \(\rho_\odot = \rho(r_\odot) = 0.4~\rm{GeV/cm^3}\)~\cite{Cirelli:2024ssz}.

\subsection{Dark matter spikes around black holes}
A further relevant consideration is the potential existence of dark matter spikes around black holes, which are localized density enhancements arising from the adiabatic growth of compact massive objects, such as supermassive black holes (SMBHs)~\cite{Gondolo:1999ef}. These spikes could substantially enhance the expected signals in indirect dark matter detection searches.

An initial power-law dark matter density profile evolves into a spike profile~\cite{Gondolo:1999ef, Quinlan:1994ed}

\begin{equation}
\rho_{\text{sp}}(r) = \rho_R\, g_\gamma(r) \left( \frac{R_{\text{sp}}}{r} \right)^{\gamma_{\text{sp}}},
\end{equation}
with the spike radius $R_{\text{sp}}$ given by

\begin{equation}
R_{\text{sp}} = \alpha_\gamma r_s \left( \frac{M_{\text{BH}}}{\rho_s r_s^3} \right)^{1/(3-\gamma)},
\end{equation}
where $\gamma_{\text{sp}} = \frac{9-2\gamma}{4-\gamma}$ is the spike power-law index, and the factors $\alpha_\gamma$ and $g_\gamma(r)$ are obtained numerically. For typical values of $\gamma = 0.7-1.4$, we have $\alpha_\gamma \simeq 0.1$ and  $g_\gamma(r) \simeq 1 - \frac{4R_\mathrm{S}}{r}$~\cite{Gondolo:1999ef}, where $R_\mathrm{S}$ is the Schwarzschild radius of the black hole. Furthermore, a normalization factor $\rho_R = \rho_0 (R_{\text{sp}}/r_0)^{-\gamma}$ is needed in order to ensure continuity with the density profile outside the spike region. The spike profile is valid only for $4R_S \leq r \leq R_\mathrm{sp}$, vanishing at smaller radii due to capture by the black hole. At radii $r > R_\mathrm{sp}$, the dark matter distribution maintains its original power-law profile (e.g. NFW).

This spike profile only holds for non-annihilating dark matter or in scenarios where annihilation is negligible, such as the DPDM case we are considering. However, for dark matter models with substantial annihilation cross-sections,  the central density saturates at an equilibrium value, leading to the formation of an inner core.

The spike normalization is determined under the constraint that the enclosed dark matter mass at radii relevant for black hole mass determinations (typically $\sim 10^5 R_S$~\cite{Neumayer_2010,2009ApJ...700.1690G}) does not exceed the uncertainty on the black hole mass $\Delta M_{\text{BH}}$~\cite{PhysRevD.82.083514},

\begin{equation}
\int_{4R_s}^{10^5 R_s} 4\pi r^2 \rho_\mathrm{DM} \, dr \lesssim \Delta M_{\text{BH}}.
\label{eq:normalization}
\end{equation}

\section{Astrophysical sources}
\label{sec:sources}

According to Eq.~\ref{eq:threshold}, photon–dark photon scattering becomes kinematically allowed at energies above $52~\mathrm{TeV}$ for dark photon masses of $m_{A^\prime} = 10^{-2}~\mathrm{eV}$. This energy threshold establishes the minimum gamma-ray energies needed to probe the model, motivating the selection of astrophysical sources with significant emission in the TeV regime. In this section, we present the selected sources, outlining their overall characteristics, spectral properties and the associated dark matter density distributions considered in the analysis.

\subsection{Crab Nebula}
\label{subsec:sources_crab}

The Crab Nebula is a galactic source located in the Taurus constellation at a distance of approximately 2 kpc, and represents one of the most important and well-studied VHE sources in astrophysics~\cite{2008ARA&A..46..127H}. This system originated from a supernova explosion observed in AD 1054 and consists of three primary components: the Crab Pulsar PSR B0531+21, the surrounding pulsar wind nebula and the expanding supernova remnant itself~\cite{2008ARA&A..46..127H, Buhler:2013zrp, Amato}. The Crab Nebula dominates as one of the brightest high-energy and very-high-energy sources in the sky. Its exceptional luminosity, coupled with remarkably stable emission properties, has established it as the standard candle for VHE astronomy and a natural laboratory for studying particle acceleration and non-thermal emission processes in extreme astrophysical environments.

The spectral energy distribution of the nebula reveals two distinct components: synchrotron emission from relativistic electrons in the nebula's magnetic field dominates the lower  energies ($\lesssim$ 1 GeV), while inverse Compton (IC) scattering of ultrarelativistic electrons produces the high-energy peak~\cite{Rees,Aharonian,Hillas,2010A&A...523A...2M,HESS:2024fri}. Recent detections of PeV photons by LHAASO~\cite{2021Sci...373..425L} have revitalized interest in hadronic contributions to the gamma-ray flux~\cite{Peng, Nie:2022cvv}, though leptonic processes remain dominant across most of the spectrum.

At VHE, this emission is typically modeled by a log-parabola function of the form

\begin{equation}
\frac{dN}{dE} = N_0 \left(\frac{E}{E_0}\right)^{-\alpha - \beta \log(E/E_0)},
\label{eq:crab_spectral}
\end{equation}
where $E_0 = 7~\mathrm{TeV}$ is the reference energy. In our simulations, we adopt the parameters as estimated by the analysis of HAWC observations~\cite{HAWC:2019xhp}: $N_0 = (2.35 \pm 0.04) \times 10^{-13}~\mathrm{TeV}^{-1}\mathrm{s}^{-1}\mathrm{cm}^{-2}$, $\alpha = 2.79 \pm 0.02$ and $\beta = 0.10 \pm 0.01$.

\begin{figure}[htbp]
  \centering
  \includegraphics[width=0.8\linewidth]{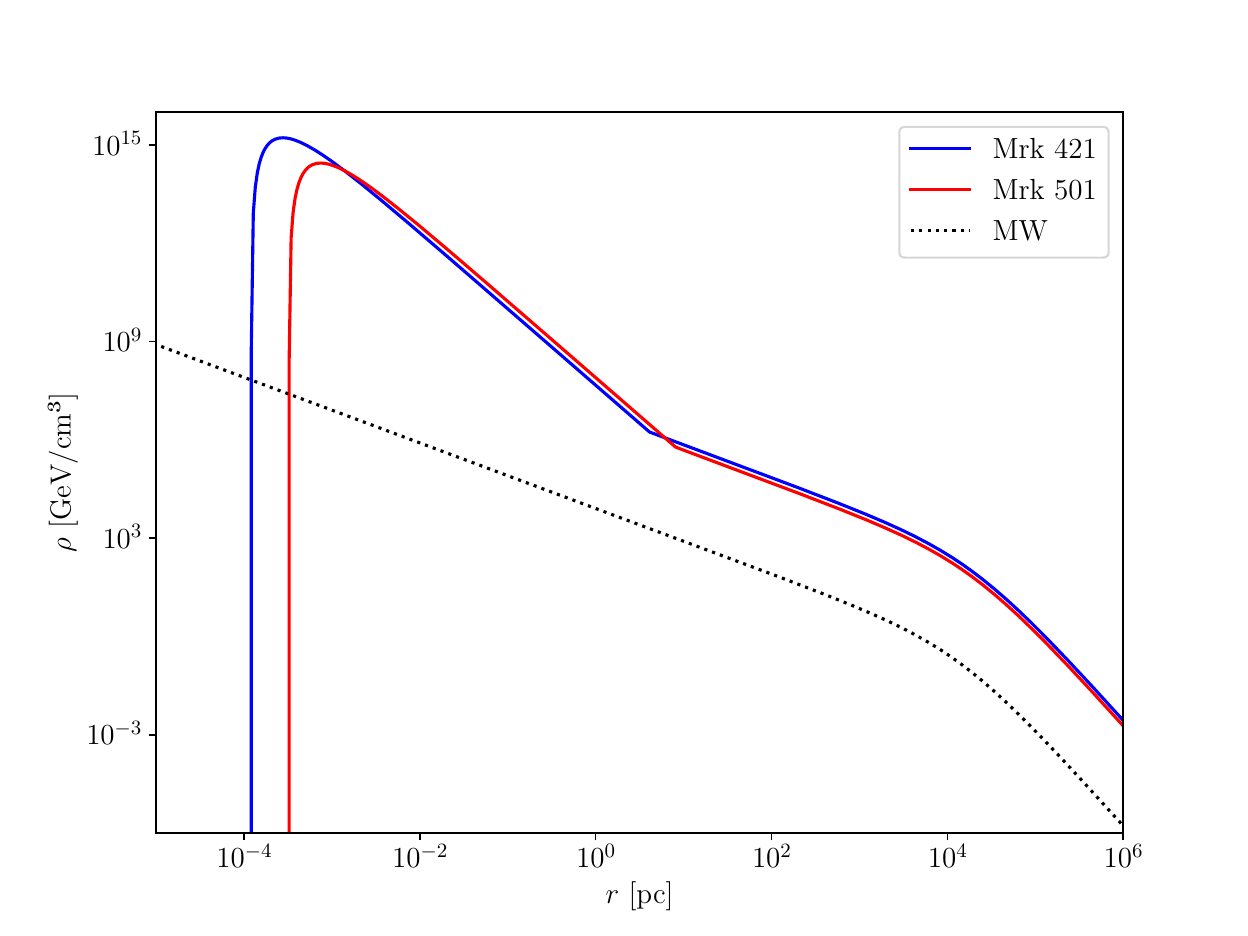}
  \caption{Radial dark matter density profiles for Mrk~421 (blue) and Mrk~501 (red) host galaxies, modeled with a dark matter spike (parameters in Table~\ref{tab:profiles}). The dotted black line shows the Milky Way's NFW profile for comparison, normalized to the local density at $r_\odot$ (black dotted line).}
  \label{fig:density_profiles}
\end{figure}

Since the Crab Nebula is a Galactic source, we model the surrounding dark matter distribution using the NFW profile (Eq.~\ref{eq:NFW}) with parameters consistent with the Milky Way halo. This profile is shown in Figure~\ref{fig:density_profiles}, and the corresponding LOS integral is computed based on the source's location, as given in Table~\ref{tab:location}. 

\subsection{Mrk 421 and Mrk 501}
\label{subsec:sources_blazars}

Markarian 421 (Mrk 421) and Markarian 501 (Mrk 501) are two of the most 
extensively studied blazars in VHE gamma-ray astronomy. These objects are classified as BL Lacertae-type blazars, a subclass of active galactic nuclei (AGN) characterized by featureless optical spectra and strong non-thermal emission. Their jets are aligned close to the observer's line of sight, leading to significant relativistic boosting and making them prominent sources in the VHE regime~\cite{Ghisellini_2009, Gaidos:1996ss,1992Natur.358..477P}.

Mrk 421, located in the constellation Ursa Major, was the first extragalactic object detected at TeV energies, discovered in 1992 by the Whipple Observatory~\cite{1992Natur.358..477P}. Mrk 421 is known for its dramatic flaring episodes in the TeV band. In 2001, it exhibited a historic flare where the flux exceeded the Crab Nebula flux~\cite{Krennrich_2002, AharonianMrk4221}. 

Mrk 501, in the constellation Hercules, became the second confirmed TeV blazar after detection in 1996 by the Whipple collaboration~\cite{Quinn1996}. During a major 1997 flare, its VHE flux exceeded that of the Crab Nebula, with its emission also being observed beyond 20 TeV~\cite{Aharonian:1999yd}. 

Such events offer critical insights into particle acceleration and radiation mechanisms in AGN jets. Simultaneous multiwavelength campaigns on both sources, using instruments such as MAGIC, HAWC, HEGRA, VERITAS, H.E.S.S., and Fermi-LAT, have enabled modeling of their spectral energy distributions (SEDs) and testing of leptonic and hadronic models~\cite{Albert_2007,AharonianMrk4221,Aharonian501,Aharonian_2009,Abdo_2011_421,Albert:2021fim,Acciari_2011_421,Acciari_2011_501}.

For the blazar benchmarks, we adopt the HEGRA parametrizations of the observed VHE spectra of Mrk~421 and Mrk~501, described by an exponential cutoff power-law (ECPL) function

\begin{equation}
\frac{dN}{dE} = N_0 \left(\frac{E}{1\,\text{TeV}}\right)^{-\alpha} \exp\left(-\frac{E}{E_\mathrm{cut}}\right),
\label{eq:spectra}
\end{equation}
where $N_0$, $\alpha$, and $E_\mathrm{cut}$ are the normalization, spectral index, and cutoff energy, respectively.

The spectral parameters adopted in this work are taken from HEGRA measurements of Mrk~421 and Mrk~501 during bright activity states. For Mrk~421, we use the spectrum measured between Nov.~2000 and May~2001, with a total exposure of $254.8~\mathrm{h}$, a period that included a historically bright state of the source. The corresponding ECPL parameters are $N_0 = (11.4 \pm 0.3)\times 10^{-11} ~\mathrm{TeV^{-1}s^{-1}cm^{-2}}$, $\alpha = 2.19 \pm 0.02$, and $E_\mathrm{cut} = 3.6~\mathrm{TeV}$~\cite{AharonianMrk4221}. For Mrk~501, we adopt the HEGRA spectrum averaged over several months of the 1997 outburst, with $N_0 = (10.8\pm 0.2)\times 10^{-11} ~\mathrm{TeV^{-1}s^{-1}cm^{-2}}$, $\alpha = 1.92 \pm 0.03$, and $E_\mathrm{cut}= 6.2\pm0.4~\mathrm{TeV}$~\cite{Aharonian:1999vy}. 

These spectra are therefore not intended to represent the long-term average emission of the blazars, but rather benchmark bright-state spectra for CTAO sensitivity projections. This choice also enables a direct comparison with previous VHE studies of the same dark photon dark matter scenario.

After specifying the source spectra, we model the dark matter distribution encountered by the photons along their path to Earth. For extragalactic sources such as Mrk~421 and Mrk~501, the attenuation induced by photon-dark photon scattering receives contributions not only from the immediate environment of the source, but also from the intergalactic medium and from the Milky Way halo. The corresponding optical depth can therefore be decomposed as

\begin{equation}
\begin{split}
\tau_{A^\prime} &= \tau_{A^\prime,\text{host}} + \tau_{A^\prime,\text{IGM}} + \tau_{A^\prime,\text{MW}} \\
&= \int_{\text{los}} (\rho_{\text{DM},\text{host}} + \rho_{\text{DM}, \text{IGM}} + \rho_{\text{DM}, \text{MW}}) \times (\sigma_{A^\prime}/m_{A^\prime})\, dl.
\end{split}
\end{equation}

For the Milky Way's dark matter density distribution, we adopt the NFW profile (Eq.~\ref{eq:NFW}) with the standard Galactic parameterization described in Section~\ref{sec:density}. The intergalactic medium (IGM) contains a very low density of dark matter, which can be approximated as 
$\rho_\mathrm{DM,IGM} = \Omega_\mathrm{DM} \times \rho_c$. 
Here, $\Omega_\mathrm{DM} \approx 0.26$ is the present-day dark matter 
density parameter, and $\rho_c \approx 8.64 \times 10^{-30} \, 
\text{g cm}^{-3}$ is the critical density of the universe in the 
$\Lambda\mathrm{CDM}$ model~\cite{Planck:2018vyg}.

For the host galaxies of the sources Mrk~421 and Mrk~501, we use a spike profile as described in Section~\ref{sec:density}, with a NFW profile adopted at radii $r>R_\mathrm{sp}$. The scale density $\rho_s$ and the spike radius $R_\mathrm{sp}$ are set by taking the upper bound in Eq.~\ref{eq:normalization}, while the scale radius is fixed as $r_s = 20\,\mathrm{kpc}$, consistent with typical parameterizations for galaxies lacking well-constrained rotation curve measurements~\cite{2010arXiv1005.0411C, Tormen:1996fc, halo_2010A&A...511A..89C, Lacroix:2016qpq}. The parameters of both host galaxies are displayed in Table~\ref{tab:profiles}, and the corresponding DM density profiles are shown in Figure~\ref{fig:density_profiles}.

\begin{table}[ht!]
    \centering
    \begin{tabular}{l c c c}
         \toprule
Source & {R.A. (hh mm ss) } & {Dec (dd mm ss)} & {Distance}  \\
\midrule
Crab & 05 34 30.9 & +22 00 44.5 & 2~kpc \\
Mrk 421 & 11 04 19 & +38 11 41 & z = 0.031 \\
Mrk 501 & 16 53 52.2  & +39 45 37 & z = 0.034 \\
\bottomrule
    \end{tabular}
    \caption{Distances and equatorial coordinates of the sources~\cite{TeVCat}.}
    \label{tab:location}
\end{table}

\begin{table}[ht!]
    \centering
    \begin{tabular}{l c c c c c}
         \toprule
Source & $\gamma$ & $\rho_s(\mathrm{GeVcm^{-3}})$ &  $r_s(\mathrm{kpc})$ &  $R_\mathrm{sp} (\mathrm{pc})$&{$\log(M_\mathrm{BH}/M_\odot)$}  \\
\midrule
Mrk 421 &1& 357.73 & 20 & 4.10 & $8.50 \pm 0.18$\\
Mrk 501 &1& 244.73 & 20 & 8.13 & $8.93 \pm 0.21$\\
\bottomrule
    \end{tabular}
    \caption{Parameters of Mrk~421 and Mrk~501 dark matter density profiles. The SMBH masses and their associated uncertainties were adopted from Ref.~\cite{Falomo}.}
    \label{tab:profiles}
\end{table}

\begin{figure}[ht!]
  \centering
  \includegraphics[width=0.8\linewidth]{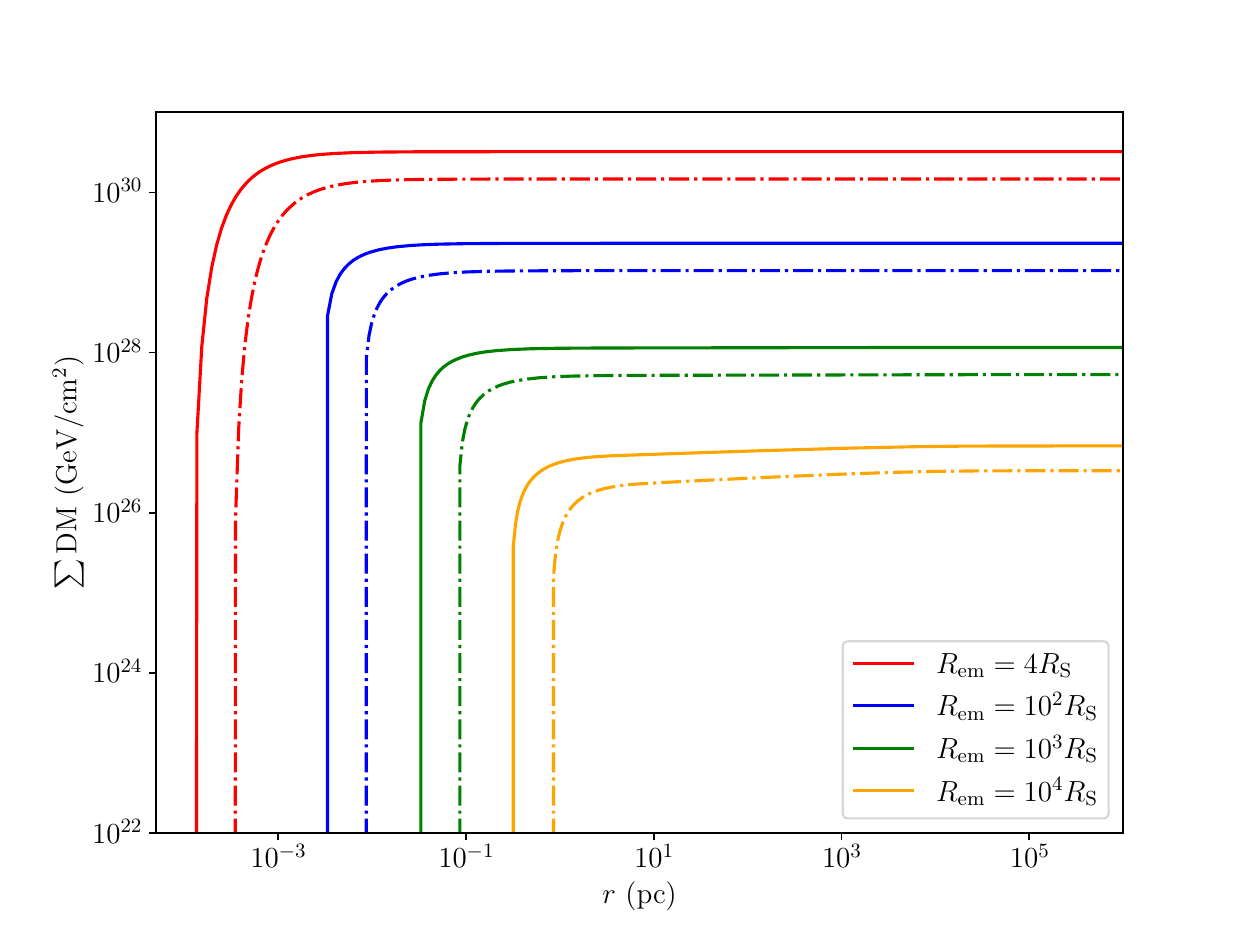}
  \caption{Line-of-sight (LOS) integrals of the dark matter mass for the host galaxies of Mrk~421 (solid lines) and Mrk~501 (dash-dotted lines), computed for different source emission regions $R_{\mathrm{em}}$, where $R_S$ denotes the Schwarzschild radius of the central SMBH.
}
  \label{fig:LOS}
\end{figure}

The VHE gamma-ray emissions are produced in compact regions situated relatively close to their central SMBHs, with radial distances that may vary in the broad range of $R_\mathrm{em}\approx (4-10^4)R_\mathrm{S}$~\cite{Abdo_2011_421,Blasi:2013tta,Abdo_2011_501,2024Univ...10..114L}.  The DM mass integrals along the line-of-sight for Mrk~421 and Mrk~501 host galaxies, considering four different emission regions $R_\mathrm{em}$, are shown in Figure~\ref{fig:LOS}. The contributions of the MW halo and IGM to the total integrated DM mass are negligible when compared to the dominant host galaxy's dark matter distribution, being smaller by a factor of $\sim 10^{-8}$ and $\sim 10^{-9}$ respectively.  

\section{CTAO simulated observations}
\label{sec:simulations}

The Cherenkov Telescope Array Observatory (CTAO) is a next-generation ground-based observatory designed to explore the VHE gamma-ray sky with unprecedented sensitivity and angular resolution. With its broad energy coverage from $20~\mathrm{GeV}$ to beyond $300~\mathrm{TeV}$, CTAO will investigate extreme astrophysical phenomena including PeVatrons, supernova remnants, and relativistic jets in active galactic nuclei~\cite{CTA-Consortium2019}. The observatory’s improved energy resolution and low energy threshold will enable landmark studies of indirect dark matter signatures. These include searches for gamma-ray lines from DM annihilation or decay in the Galactic Center, dwarf spheroidal galaxies and galaxy clusters, as well as broadband emission from TeV-scale WIMP's (Weakly Interacting Massive Particles)~\cite{Viana:2019ucn,CTAO:2024wvb,Duangchan:2022jqn}. Furthermore, CTAO’s large effective area and ability to resolve spectral cutoffs will also test DM particle models beyond the standard WIMP paradigm, such as dark photons and axion-like particles~\cite{Liang:2018mqm,Reis:2024wfy,linden2024indirectsearchesdarkphotonphoton}. 

This section details our simulation framework for spectral analysis of the three target sources, using publicly available instrument response functions (IRFs) from the CTA Prod5 dataset. Our analysis pipeline is implemented with \textsc{Gammapy} version 1.3~\cite{gammapy:2023,acero_2025_14760974}, and includes data simulation, instrument response modeling, and spectral extraction for each source. For all sources, we perform a one-dimensional ON--OFF analysis. The ON region is defined as a circular region of radius $0.5^\circ$ centered on the target position. Since the sources considered here are treated as point-like in the CTAO simulations, this aperture is chosen to contain the source emission over the energy range used in the analysis. The residual background is estimated using the reflected-region method: for each pointing, ten OFF regions with the same angular radius as the ON region are placed at the same offset from the pointing direction and at different azimuthal positions. This configuration provides OFF regions with approximately the same instrumental acceptance as the ON region. The corresponding ON/OFF exposure ratio is therefore $\alpha_{\rm on/off} \simeq 1/N_{\rm off}=0.1$.

\subsection{Crab Nebula spectral simulations}
\label{subsec:crab_simu}

Simulated observations of the Crab Nebula were performed using the Northern array layout of the CTAO Alpha Configuration, consisting of 4 Large-Sized Telescopes (LSTs) and 9 Medium-Sized Telescopes (MSTs). We used the instrument response functions (IRFs) from the Prod5 dataset, corresponding to the ``North-20deg-AverageAz'' configuration\footnote{
The IRFs employed in this analysis correspond to a zenith angle of $20^\circ$. The effect of adopting larger zenith angles for the Northern array was assessed separately and found not to significantly modify the derived dark photon constraints within the mass range considered in this work.
}, optimized for 50 hours of observation time~\cite{cherenkov_telescope_array_observatory_2021_5499840}, which include the effective area, energy dispersion, and point spread function. A wobble observation strategy was adopted, with four offset pointings located $0.4\degree$ from the target position, at position angles of $0\degree$, $90\degree$, $180\degree$, and $270\degree$. 

The total simulated livetime of 200 hours was evenly distributed across these wobble positions. This exposure reflects the role of the Crab Nebula as a standard candle and calibration source in very-high-energy gamma-ray astronomy, for which accumulated observation times of this order are realistic within the CTAO lifetime. The adopted exposure is intended to provide a representative benchmark scenario rather than an optimized observing strategy.

The simulation covered a reconstructed energy range from 0.05 to 200\,TeV, with 20 bins per decade, and a true energy range from 0.025 to 300\,TeV, with 40 bins per decade, in order to accurately model the instrument response functions and account for IRF tails. A safe energy threshold was defined using \textsc{Gammapy}’s \texttt{SafeMaskMaker} with the \texttt{edisp-bias} method, retaining only energy bins where the energy-dispersion bias, as derived from the instrument response functions (IRFs), is below 10\%. This defines a safe reconstructed-energy range consistent with CTAO real-data analyses, ensuring reliable flux reconstruction and limiting systematic effects from the energy dispersion. The common ON--OFF configuration described above was then applied to construct the one-dimensional spectral dataset.

\begin{figure}[ht!]
  \centering
  \includegraphics[width=0.8\linewidth]{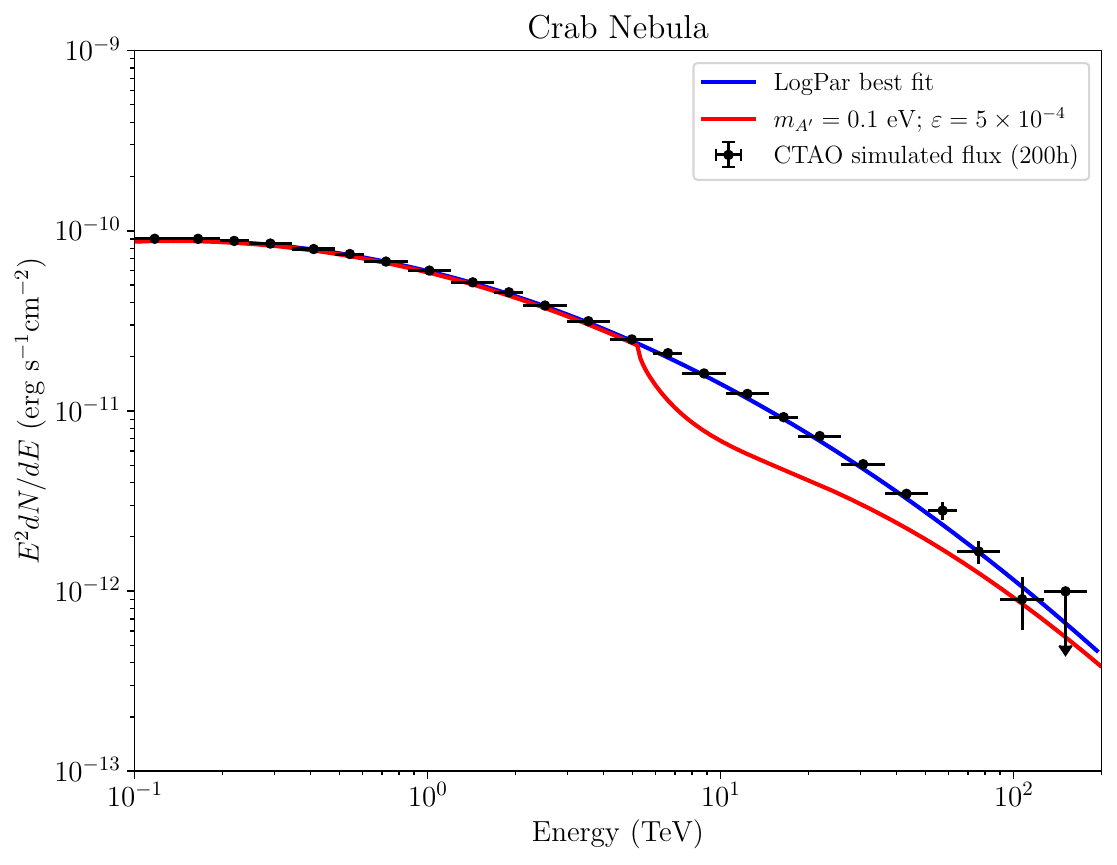}
  \caption{The black points show one simulated CTAO flux realization for the Crab Nebula as $E^2 dN/dE$ estimated in logarithmically spaced energy bins.
 The blue line represents the best-fit log-parabola ($H_0$), while the red line illustrates the dark photon scattering attenuation effect for $m_{A^\prime} = 0.1~\mathrm{eV}$ and $\varepsilon = 5\times10^{-4}$. The best-fit parameters for the null hypothesis are shown in Table~\ref{tab:bestfit}.}
  \label{fig:FluxCrab}
\end{figure}

The Crab Nebula's VHE emission was modeled as a point source centered at the coordinates shown in Table~\ref{tab:location}, with a log-parabolic spectral energy distribution (Eq.~\ref{eq:crab_spectral}), using the parameters detailed in Section~\ref{subsec:sources_crab}. The resulting simulated differential flux points are shown in Figure~\ref{fig:FluxCrab}, as well as the assumed SED model, with and without DPDM attenuation. 

\subsection{Mrk 421 and Mrk 501 spectral simulations}

Simulated observations of the blazars Mrk~421 and Mrk~501 were also performed using the Northern array layout of the CTAO Alpha Configuration, consisting of 4 Large-Sized Telescopes (LSTs) and 9 Medium-Sized Telescopes (MSTs). The instrument response functions (IRFs) were taken from the Prod5 dataset, corresponding to the ``North-20deg-AverageAz'' configuration, optimized for 50 hours of observation time~\cite{cherenkov_telescope_array_observatory_2021_5499840}. The same wobble observation strategy used for the Crab Nebula was adopted.

For each blazar, we assume a total simulated livetime of 100 hours, equally distributed across the wobble positions. This exposure is intended to represent a realistic dedicated observing campaign for bright variable blazars. Although such observation times exceed those foreseen for the early CTAO phases, comparable cumulative exposures are expected to be achieved as the observatory approaches its full operational configuration. Moreover, adopting 100 hours per source enables meaningful comparison with previous VHE studies and commonly adopted sensitivity references in the literature.

The reconstructed energy range spanned from 0.05 to 50\,TeV, with 20 bins per decade, while the true energy extended from 0.025 to 100\,TeV, sampled using 40 bins per decade to ensure proper resolution of instrument response effects.
The same safe-energy selection based on the 10\% energy-dispersion bias criterion described in Section~\ref{subsec:crab_simu} was applied to the blazar datasets, together with the ON--OFF configuration and reflected-region background treatment described above.

\begin{figure}[ht!]
  \centering
  \includegraphics[width=0.8\linewidth]{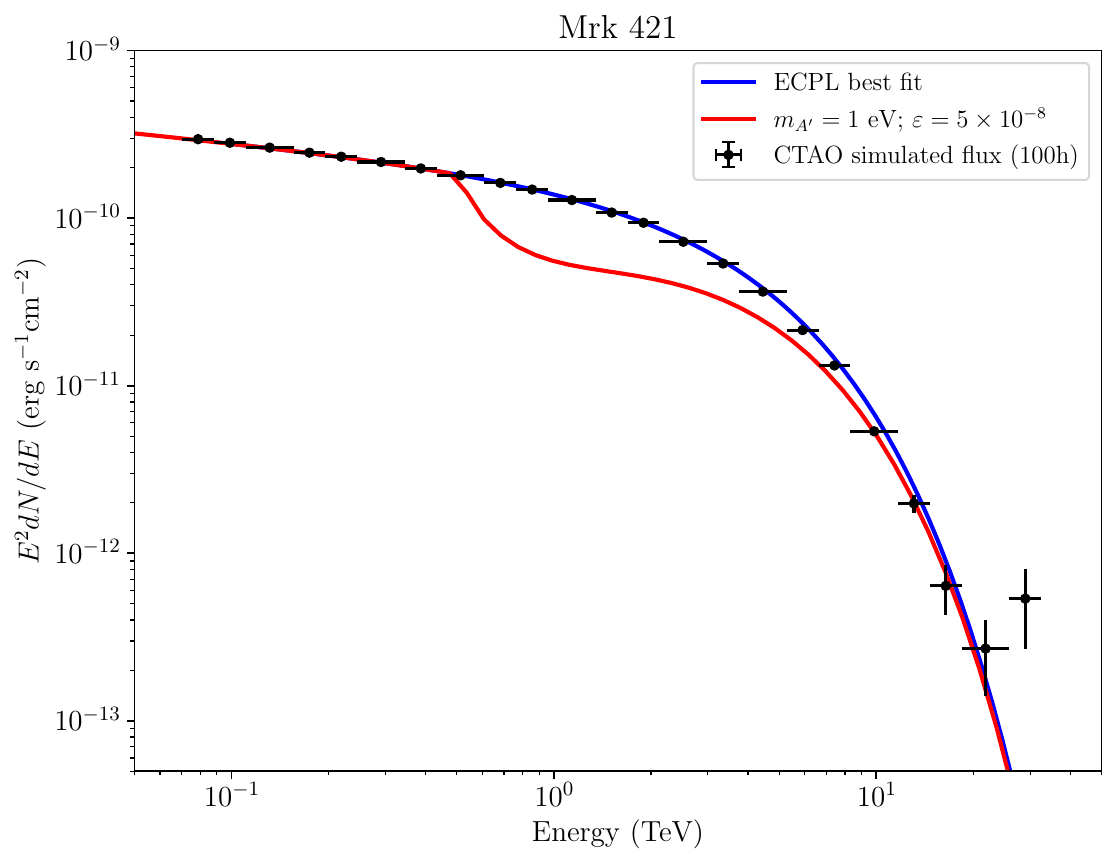}
  \caption{The black points show one simulated CTAO flux realization for Mrk~421 as $E^2dN/dE$, estimated in logarithmically spaced energy bins. The blue line represents the best-fit ECPL function ($H_0$), while the red line illustrates the dark photon scattering attenuation effect for $m_{A^\prime} = 1~\mathrm{eV}$ and $\varepsilon = 5\times10^{-8}$. The best-fit parameters for the null hypothesis are shown in Table~\ref{tab:bestfit}.}
  \label{fig:FluxMrk421}
\end{figure}

\begin{figure}[ht!]
  \centering
  \includegraphics[width=0.8\linewidth]{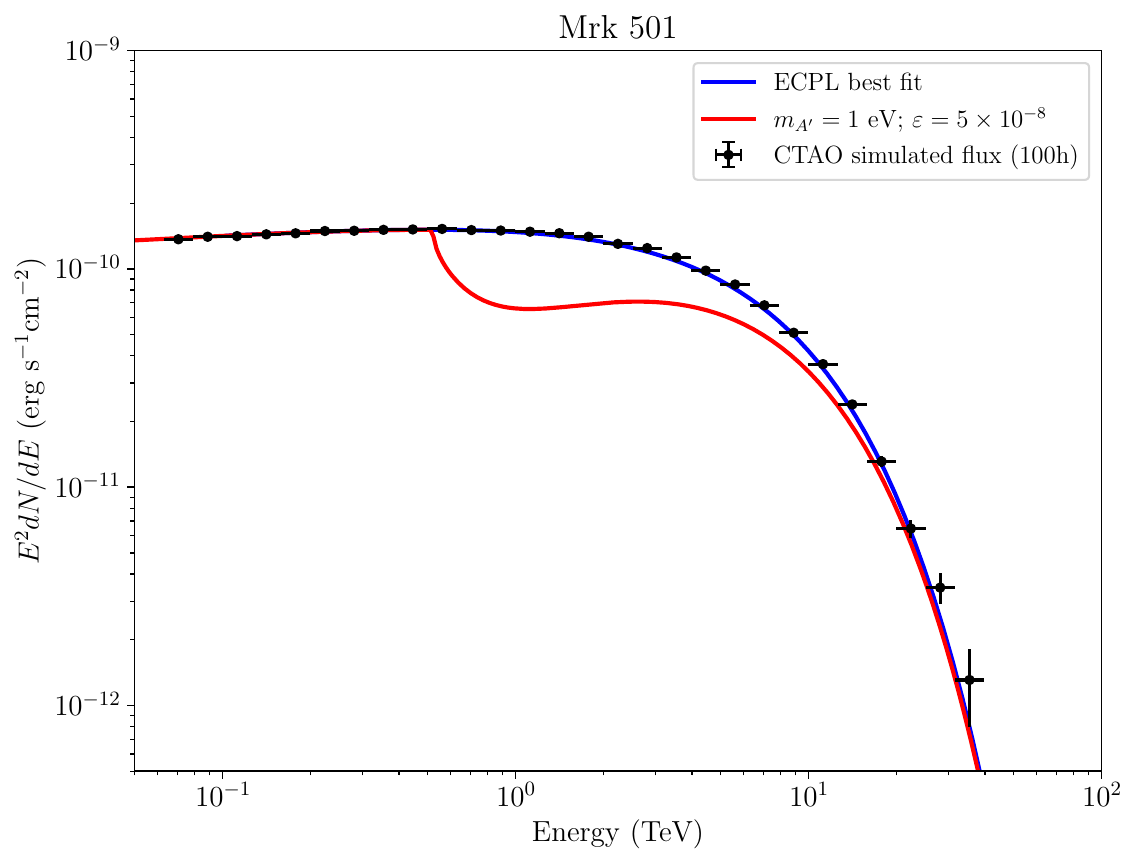}
  \caption{The black points show one simulated CTAO flux realization for Mrk~501 as $E^2 dN/dE$ estimated in logarithmically spaced energy bins. The blue line represents the best-fit ECPL function ($H_0$), while the red line illustrates the dark photon scattering attenuation effect for $m_{A^\prime} = 1~\mathrm{eV}$ and $\varepsilon = 5\times10^{-8}$. The best-fit parameters for the null hypothesis are shown in Table~\ref{tab:bestfit}.}
  \label{fig:FluxMrk501}
\end{figure}

Mrk~421 and Mrk~501 were modeled as point sources centered at the coordinates
shown in Table~\ref{tab:location}, with ECPL spectra (Eq.~\ref{eq:spectra}), using the spectral parameters described
in Section~\ref{subsec:sources_blazars}. Since these HEGRA parametrizations correspond to observed VHE spectra, the EBL attenuation is already effectively included in the benchmark spectral shapes, and no additional EBL attenuation is applied in the forward simulations. The resulting simulated differential flux points are shown in Figures~\ref{fig:FluxMrk421} and~\ref{fig:FluxMrk501}.

\section{Model fitting}
\label{sec:fit}

In this section, we describe the procedure adopted to determine the best-fit parameters of the spectral models and to derive 95\% confidence level upper limits on the DPDM kinetic mixing parameter $\varepsilon$ using a likelihood-based statistical approach.

\subsection{Null Hypothesis $H_0$}
\label{subsec:H0}

For the null hypothesis $H_0$, we consider the scenario in which the gamma-ray flux is fully explained by standard astrophysical processes at the source, such as particle acceleration and diffusion, together with the standard propagation effects relevant for each target. As such, $H_0$ defines the baseline physical scenario in the absence of photon-dark photon interactions and is used to generate the simulated datasets employed in the Monte Carlo realizations.

In general, the observed spectrum under the null hypothesis can be written as
\begin{equation}
    \left.\frac{dN}{dE}\right|_{H_0}
    =
    \left.\frac{dN}{dE}\right|_{\mathrm{int}}
    \times f_{\mathrm{att}}(E),
    \label{eq:null}
\end{equation}
where $\left.dN/dE\right|_{\mathrm{int}}$ is the intrinsic source spectrum and $f_{\mathrm{att}}(E)$ accounts for standard propagation attenuation.

For the Crab Nebula, the effects due to Galactic propagation (e.g., $\gamma-\gamma$ interactions with the interstellar radiation field (ISRF) or cosmic microwave background (CMB)) are typically negligible below $\sim100~\mathrm{TeV}$~\cite{2021Natur.594...33C, Moskalenko:2005ng}. Therefore, we assume $f_{\mathrm{att}}(E) \approx 1$.

For Mrk~421 and Mrk~501, the HEGRA ECPL parametrizations adopted in Section~\ref{subsec:sources_blazars} correspond to observed VHE spectra. Therefore, in the practical implementation of the null hypothesis, these spectra are used directly as empirical no-DPDM benchmarks at Earth,
\begin{equation}
    \left.\frac{dN}{dE}\right|_{H_0}^{\mathrm{blazar}}
    =
    \left.\frac{dN}{dE}\right|_{\mathrm{HEGRA,obs}} .
    \label{eq:null_blazar}
\end{equation}
In this formulation, the standard propagation effects already encoded in the observed HEGRA spectra are absorbed into the empirical benchmark used for the forecast.

\subsection{Alternative Hypothesis $H_1$}
\label{subsec:H1}

The alternative hypothesis $H_1$ extends the baseline spectral model by including the photon survival probability induced by photon-dark photon scattering, given by Eq.~\ref{eq:survival}. The resulting spectral model is

\begin{equation}
    \left.\frac{dN}{dE}\right|_{H_1}
    =
    P_\mathrm{sur}(m_{A^\prime};\varepsilon)
    \times
    \left.\frac{dN}{dE}\right|_{H_0},
    \label{eq:alternative}
\end{equation}
where the dependence on the dark photon parameters $m_{A^\prime}$ and $\varepsilon$ arises from the scattering cross-section (Eq.~\ref{eq:xsec}).

This construction provides a practical way to test for an additional DPDM-induced attenuation on top of the baseline spectrum, but it relies on an assumption about the intrinsic source emission. In particular, we assume that the same baseline spectral shape used under \(H_0\) can be used under \(H_1\), with the DPDM contribution entering only through the multiplicative survival probability \(P_{\rm sur}\). For the blazar targets, this means that possible intrinsic spectral structures that could mimic or compensate for a DPDM-induced attenuation are not modeled independently. The projected limits derived here are therefore conditional on this baseline-spectrum assumption.

\subsection{Log-likelihood statistical test}

To derive constraints on the DPDM parameters, we adopt a profile-likelihood ratio approach based on the Likelihood Ratio Test (LRT) framework. The likelihood is evaluated using the W-statistic (WStat)~\cite{Cash}, as implemented in Gammapy~\cite{gammapy:2023}. This statistic is appropriate for ON--OFF spectral measurements with Poisson-distributed counts, and accounts for the statistical uncertainty of the background estimate through the OFF counts. The WStat fit statistic per reconstructed-energy bin is given by~\cite{gammapy:2023}

\begin{equation}
W = 2 \big(\mu_{\mathrm{sig}} + (1 + 1/\alpha_{\rm on/off})\mu_{\mathrm{bkg}}
- n_{\mathrm{on}} \log{(\mu_{\mathrm{sig}} + \mu_{\mathrm{bkg}})}
- n_{\mathrm{off}} \log{(\mu_{\mathrm{bkg}}/\alpha_{\rm{on/off}})}\big),
\end{equation}
where $n_\mathrm{on}$ and $n_\mathrm{off}$ are the observed counts in the ON and OFF regions, respectively. The quantity $\mu_\mathrm{sig}$ is the expected number of source counts in the ON region, obtained by folding the spectral model with the CTAO instrument response functions, while $\mu_\mathrm{bkg}$ is the expected residual background contribution in the ON region. The parameter $\alpha_{\rm on/off}$ is the exposure ratio between the ON and the combined OFF regions. For our reflected-region configuration with ten OFF regions of equal size and approximately equal acceptance, $\alpha_{\rm on/off} \simeq 1/10 = 0.1$.


Upper limits on the kinetic mixing parameter are derived using the profile-likelihood test statistic
\begin{equation}
TS(m_{A^\prime}; \varepsilon) =
W_{H_1}(m_{A^\prime}; \varepsilon)
-
W_{H_1}(m_{A^\prime}; \hat{\varepsilon}) ,
\label{eq:TS}
\end{equation}
where \(\hat{\varepsilon}\) is the value of the mixing parameter that minimizes  
the WStat under the alternative hypothesis.  

In the asymptotic regime, the test statistic is expected to follow a $\chi^2$ distribution with one degree of freedom, according to Wilks' theorem. We adopt this asymptotic approximation throughout the forecast analysis. Under this assumption, a value of $TS = 2.71$ corresponds to a one-sided 95\% confidence level upper limit on $\varepsilon$.

When fitting the Crab Nebula data, the spectral parameters \((N_0, \alpha, \beta)\) and the background normalization were left free to vary. For the DPDM contribution, we varied the mixing parameter \(\varepsilon\) while keeping the dark photon mass \(m_{A^\prime}\) fixed at each evaluation point. The same fitting procedure was applied to Mrk~421 and Mrk~501, whose free spectral parameters were \((N_0, \alpha, E_\mathrm{cut})\).

\section{Results}
\label{sec:results}

\begin{figure}[ht!]
  \centering
  \includegraphics[width=0.8\linewidth]{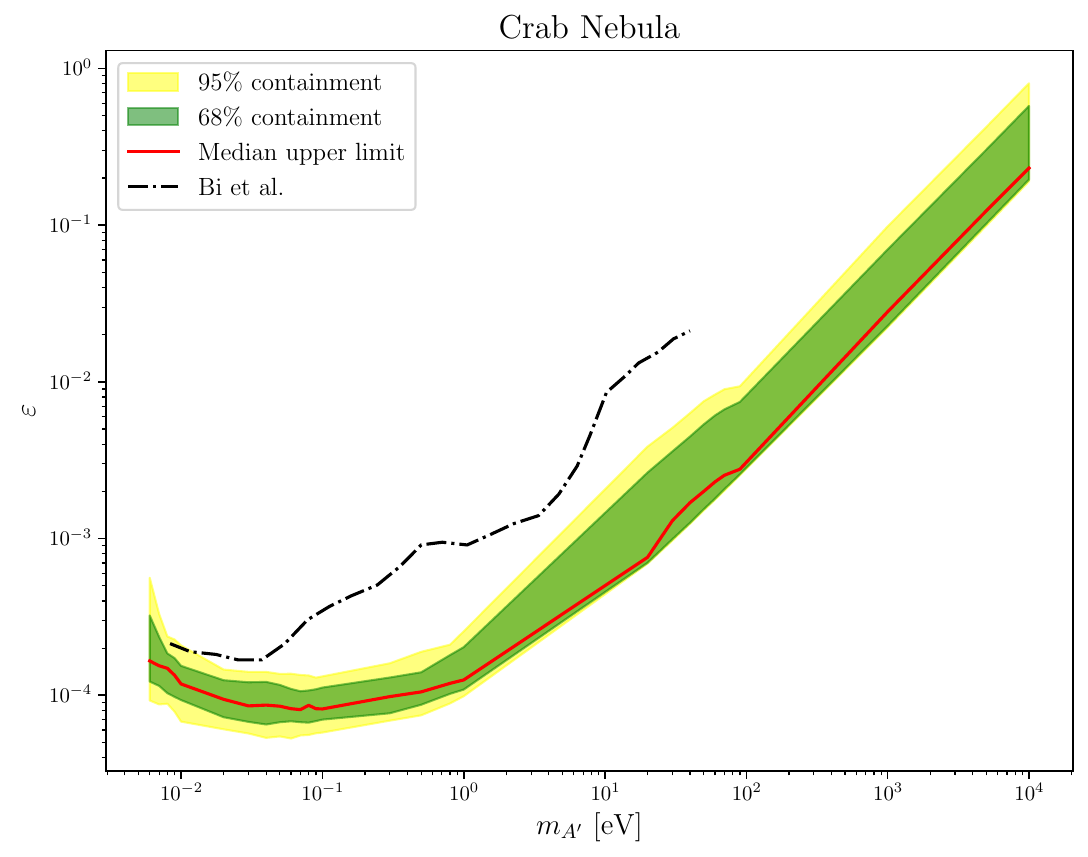}
  \caption{Expected 95\% confidence level upper limits on the kinetic mixing parameter $\varepsilon$ derived from simulated CTAO observations of the Crab Nebula. The solid red line shows the median expected upper limit, while the shaded regions indicate the 68\% and 95\% containment bands around the median. Constraints from Ref.~\cite{PhysRevD.103.043018} (dash-dotted lines) are shown for comparison.}

\label{fig:LimitsCrab}
\end{figure}
\begin{figure}[ht!]
  \centering
  \includegraphics[width=0.8\linewidth]{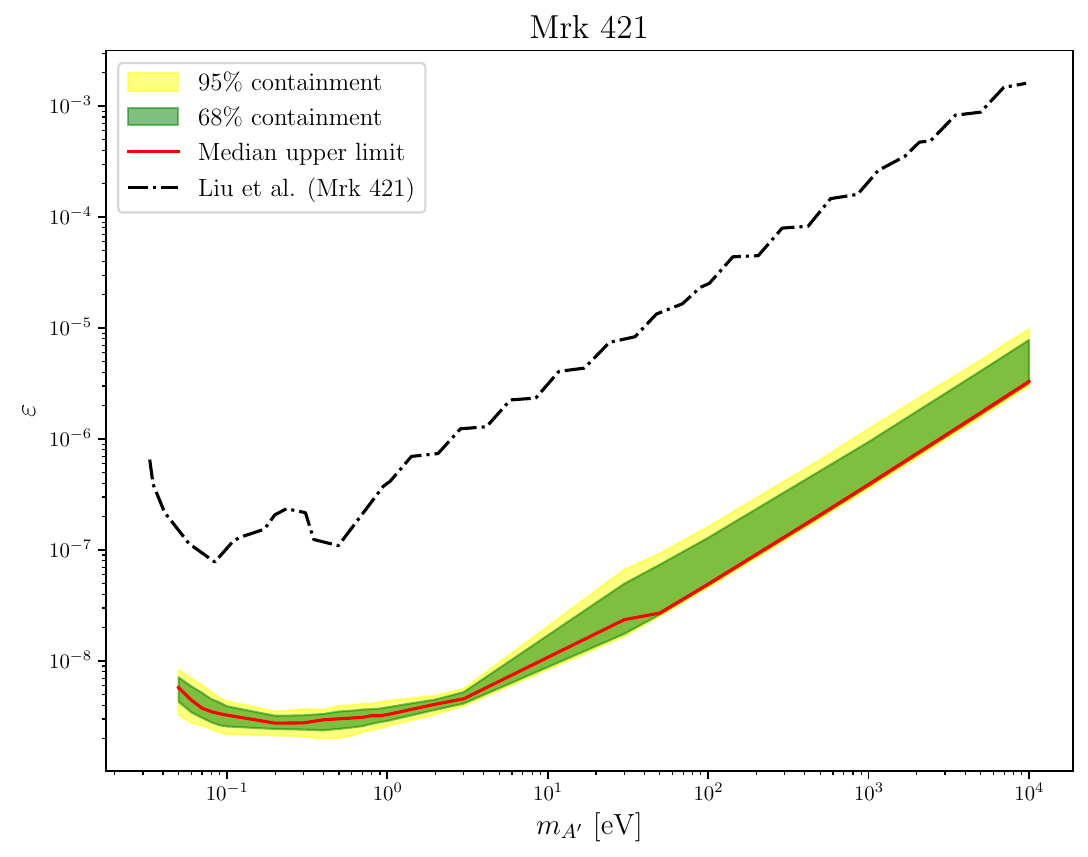}
  \caption{Expected 95\% confidence level upper limits on the kinetic mixing parameter $\varepsilon$ derived from simulated CTAO observations of Mrk~421. The solid red line shows the median expected upper limit, while the shaded regions indicate the 68\% and 95\% containment bands around the median. Constraints from Ref.~\cite{Liu2024DarkPhoton} (dash-dotted lines) are shown for comparison.}

  \label{fig:LimitsMrk421}
\end{figure}
\begin{figure}[ht!]
  \centering
  \includegraphics[width=0.8\linewidth]{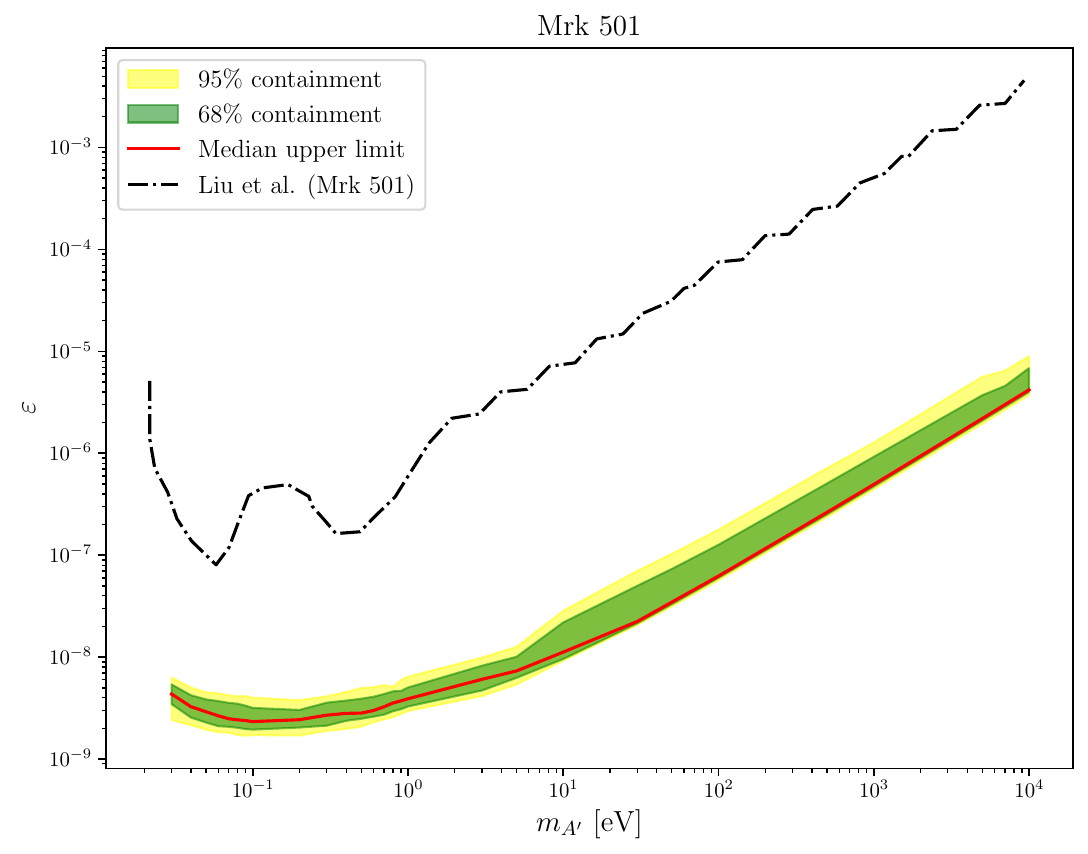}
  \caption{Expected 95\% confidence level upper limits on the kinetic mixing parameter $\varepsilon$ derived from simulated CTAO observations of Mrk~501. The solid red line shows the median expected upper limit, while the shaded regions indicate the 68\% and 95\% containment bands around the median. Constraints from Ref.~\cite{Liu2024DarkPhoton} (dash-dotted lines) are shown for comparison.}

  \label{fig:LimitsMrk501}
\end{figure}

Figures~\ref{fig:LimitsCrab}, \ref{fig:LimitsMrk421}, and~\ref{fig:LimitsMrk501} show the expected 95\% confidence level upper limits on the kinetic mixing parameter $\varepsilon$ as a function of the DPDM mass $m_{A^\prime}$, obtained for the Crab Nebula, Mrk~421, and Mrk~501, respectively. In each case, the solid line shows the median expected upper limit, and the shaded bands represent the corresponding 68\% and 95\% containment intervals obtained from 100 independent Poisson realizations of the simulated observations.

In the case of the blazars, the emission is assumed to be located at $R_{\rm em} = 4R_S$. The parameters of the best-fit spectral models used in the likelihood analysis are listed in Table~\ref{tab:bestfit}. These were obtained by fitting the simulated spectra under the null hypothesis, i.e., assuming no attenuation from photon-dark photon scattering, and define the baseline spectral models used in the subsequent likelihood scans over the DPDM mixing parameter.

Numerically, the most stringent upper limits were obtained for the blazars Mrk~421 and Mrk~501, where values as low as $\varepsilon \sim 10^{-8}$ could be constrained at the 95\% confidence level for dark photon masses around $m_{A^\prime} \sim 0.1$--$1~\mathrm{eV}$. For the Crab Nebula, which lies within the Milky Way halo and does not benefit from a central dark matter spike, the limits are comparatively weaker, constraining values of $\varepsilon \gtrsim 2\times10^{-4}$ in the $m_{A^\prime} \sim 0.01$--$0.1~\mathrm{eV}$ mass range.

\begin{figure}[ht!]
  \centering
  \includegraphics[width=0.6\linewidth]{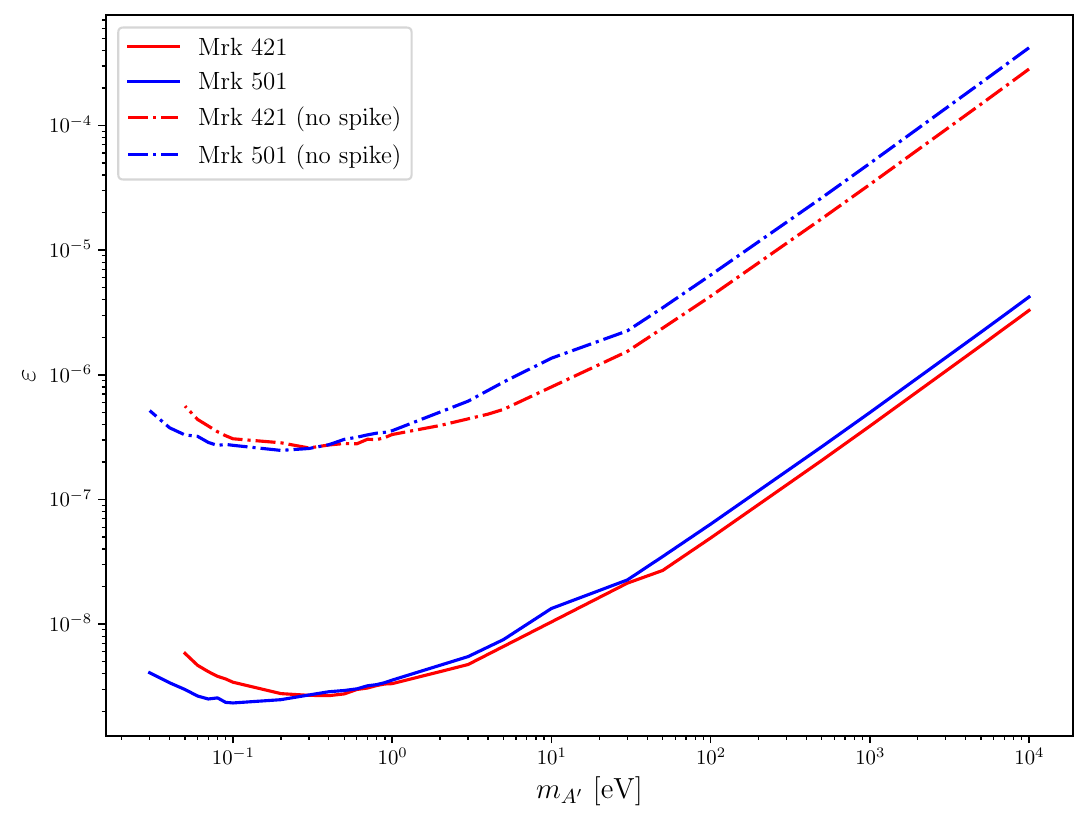}
  \caption{Expected 95\% confidence level upper limits on the kinetic mixing parameter $\varepsilon$ derived from simulated CTAO observations of Mrk~421 and Mrk~501. Solid lines correspond to scenarios including a central dark matter spike, assuming an emission region located at $R_\textrm{em} = 4R_\textrm{S}$. Dash-dotted lines show the results obtained under a canonical NFW profile with scale parameters $\rho_s$ and $r_s$ as given in Table~\ref{tab:profiles}, also assuming $R_\textrm{em} = 4R_\textrm{S}$.
}

  \label{fig:NFW_no_spike}
\end{figure}

\begin{figure}[ht!]
  \centering
  \includegraphics[width=0.6\linewidth]{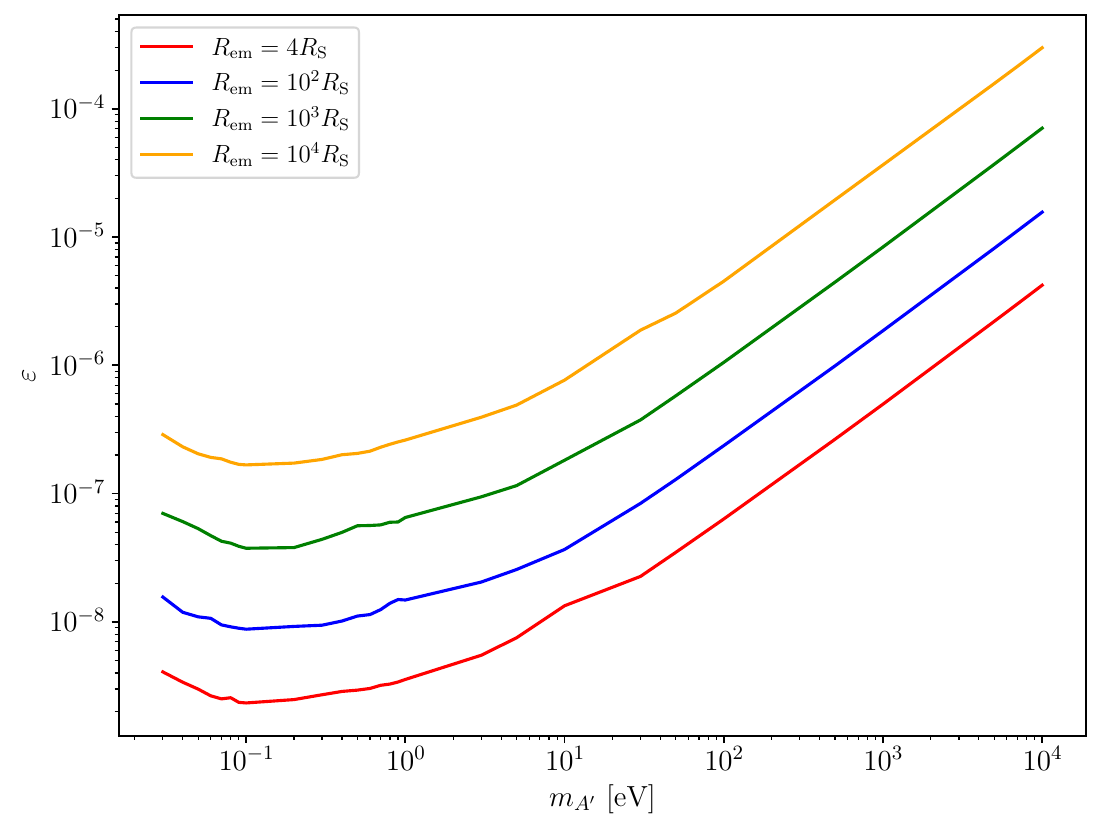}
  \caption{Expected 95\% confidence level upper limits on the kinetic mixing parameter $\varepsilon$ derived from simulated CTAO observations of Mrk~501 for different emission regions $R_\textrm{em}$.}
  \label{fig:EmissionRegions}
\end{figure}

\begin{figure}[ht!]
  \centering
  \includegraphics[width=0.8\linewidth]{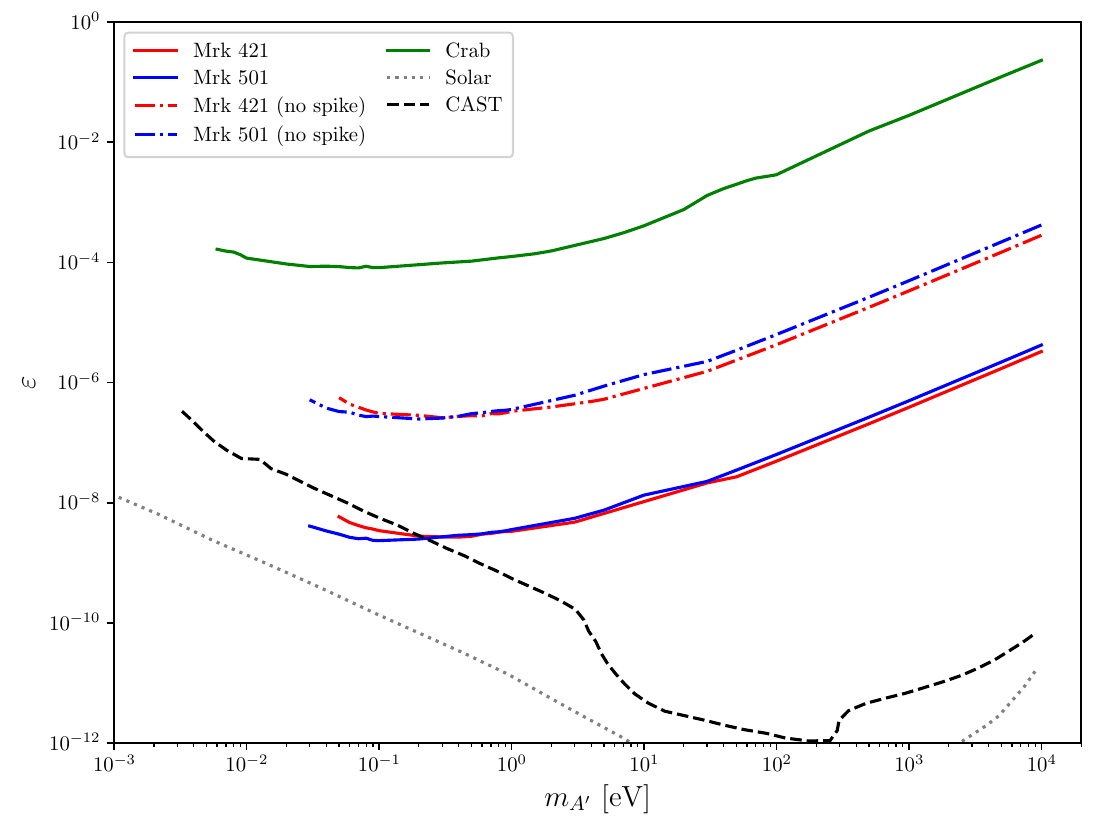}
  \caption{Constraints on the kinetic mixing parameter $\varepsilon$ derived from simulated CTAO observations of the Crab Nebula (green), Mrk~421 (red), and Mrk~501 (blue), represented as solid lines (this work). Existing bounds from solar energy loss (gray dotted line)~\cite{solar} and from the CAST helioscope experiment (black dashed line)~\cite{Redondo:2008aa} are included for comparison.}
  \label{fig:ComparedLimits}
\end{figure}

Since the analysis assumes specific dark matter distributions for each system, it is important to highlight the astrophysical uncertainties that affect the results. In particular, for Mrk~421 and Mrk~501, the presence of a dark matter spike around the central supermassive black hole plays a critical role in enhancing the optical depth to photon–dark photon scattering. However, the formation and long-term stability of such spikes are still debated, as they can be disrupted by black hole mergers or dynamical heating processes~\cite{Gondolo:1999ef,Merritt}. Figure~\ref{fig:NFW_no_spike} illustrates the impact on the constraints when the central dark matter spike is omitted and a canonical NFW profile is adopted, with $r_s$ and $\rho_s$ set to the same values as in the corresponding spiked profile scenario.

The projected sensitivities also rely on the assumption that the source spectral curvature is adequately described by the adopted empirical parametrizations. More flexible spectral models could partially absorb broad attenuation-like features and therefore weaken the projected limits. The results should thus be interpreted as benchmark CTAO sensitivity projections under the spectral assumptions described above.

The variation of the emission region’s location relative to the black hole also impacts the total traversed dark matter density, as shown in Figure~\ref{fig:LOS}. The resulting constraints on $\varepsilon$ for Mrk~501, evaluated at different values of $R_{\rm em}$, are presented in Figure~\ref{fig:EmissionRegions}. As expected, when the emission region is located far from the center, such that it lies partially or entirely outside the high-density region associated with the spike, the effective optical depth decreases. In these cases, the resulting constraints become comparable to those obtained under a canonical NFW profile without a spike.

Figure~\ref{fig:ComparedLimits} presents a combined view of the constraints derived from the three sources considered in this work, also showing the constraints from the CAST experiment~\cite{Redondo:2008aa} and from solar cooling rates~\cite{solar}. The constraints obtained with CTAO simulations for the blazars approach the sensitivity of CAST in the mass range $m_{A^\prime} \sim 0.1$--$1~\textrm{eV}$, but remain significantly weaker than those derived from solar physics. Nevertheless, the constraints derived in this work are complementary to those from CAST and solar energy loss. While the latter probe photon–dark photon mixing in laboratory settings and stellar interiors, respectively, our analysis targets a distinct physical regime, providing an independent and high-energy test of the DPDM scenario.

\begin{table}[ht]
\centering
\begin{tabular}{l c c c c}
\hline
Source & $N_0$ & $\alpha$ & $\beta$ & $E_\mathrm{cut}$ \\
& $(\text{TeV}^{-1}\text{cm}^{-2}\text{s}^{-1})$ & & &(TeV) \\
\hline
Crab Nebula & $(2.34\pm0.02)\times 10^{-13}$ & $2.79\pm0.02$ & $(1.00\pm0.03)\times10^{-1}$ &\multicolumn{1}{c}{\raisebox{0.6ex}{\rule{1cm}{0.4pt}}}\\

Mrk~421 & $(1.16\pm0.01)\times 10^{-10}$ & $2.19\pm0.01$ & \multicolumn{1}{c}{\raisebox{0.6ex}{\rule{1cm}{0.4pt}}}&$3.42\pm0.10$ \\

Mrk~501 & $(1.08\pm0.01)\times 10^{-10}$ & $1.91\pm0.01$ & \multicolumn{1}{c}{\raisebox{0.6ex}{\rule{1cm}{0.4pt}}}&$6.17\pm0.10$ \\
\hline
\end{tabular}

\caption{Best-fit parameters of the baseline spectral models used to describe the simulated CTAO observations for each source. The fits were performed under the null hypothesis, assuming no attenuation from dark photon scattering. A log-parabola spectral shape was adopted for the Crab Nebula, while an exponential cutoff power-law model was used for Mrk~421 and Mrk~501.}
\label{tab:bestfit}
\end{table}


\newpage 

\section{Conclusion}
\label{sec:conclusion}

```latex
In this work, we investigated the possibility of probing dark photon dark matter (DPDM) through spectral attenuation effects in very-high-energy (VHE) gamma rays. Using simulated CTAO observations, we derived the first dedicated CTAO forecast sensitivity to photon-dark photon scattering attenuation for a range of DPDM masses.

We considered three representative VHE sources -- the Crab Nebula, Mrk~421, and Mrk~501 -- and modeled the expected attenuation assuming dark photons make up the entire dark matter content. The resulting constraints depend on the dark matter distribution along the line of sight. In particular, for the blazars Mrk~421 and Mrk~501, the presence of a dark matter spike around the central supermassive black hole significantly enhances the predicted attenuation signal. 

Additionally, we explored the dependence of the constraints on the assumed location of the gamma-ray emission region. In compact emission scenarios, the traversed dark matter column density, and hence the optical depth, is substantially increased, leading to stronger constraints on $\varepsilon$.

While the bounds obtained here are not yet as strong as those derived from solar energy loss or laboratory experiments like CAST, they are complementary since our analysis provides an independent high-energy test of the DPDM scenario. These results illustrate CTAO's potential to contribute to the broader experimental effort to constrain hidden-sector dark matter models.

Further improvements could be achieved by including a larger sample of VHE sources with diverse spectral shapes and redshifts, as well as by refining the modeling of dark matter distributions through high-resolution simulations and multi-wavelength observations. For the blazar targets, an improved characterization of the intrinsic source spectra would also help assess the robustness of the projected limits, which assume that the baseline spectral shape adopted under $H_0$ remains valid when testing the DPDM hypothesis, with the new-physics contribution entering only as an additional attenuation factor. In addition, applying this analysis to real CTAO data in the future, once available, will allow for tests of these scenarios under more realistic observational conditions.

\clearpage
\section*{Acknowledgments}

This work was conducted in the context of the CTAO Consortium’s Dark Matter and Exotic
Physics Working Group. This research has made use of the CTA instrument response functions provided by the CTA Consortium and Observatory, see https://www.ctao.org/for-scientists/performance/ (version prod5 v0.1;~\cite{cherenkov_telescope_array_observatory_2021_5499840}) for more details.

The authors acknowledge support from the São Paulo Research Foundation (FAPESP)
under process number 2021/01089-1. VdS acknowledges support from CNPq under grant
number 308837/2023-1. JGM acknowledges support from FAPESP under process number
2022/16842-0. A.V. acknowledges support from FAPESP under process number 2024/15560-6
and from CNPq under grant number 309613/2025-6. The authors acknowledge the National Laboratory for Scientific Computing (LNCC/MCTI,  Brazil) for providing HPC resources for the SDumont supercomputer (http://sdumont.lncc.br). 

\bibliographystyle{plainnat}
\bibliography{references}



\end{document}